\documentclass[JIP]{ipsj}

\usepackage{amsmath}
\usepackage{graphicx}
\usepackage{latexsym}
\usepackage{cite}
\usepackage{url}

\usepackage{framed}

\usepackage{color}
\definecolor{darkblue}{rgb}{0.0,0,0.5} % R(赤),G(緑),B(青)
\definecolor{byzantium}{rgb}{0.44, 0.16, 0.39}

\newcommand{\nt}[1]{{\tt {\small {\color{byzantium} #1}}}}

\def\Underline{\setbox0\hbox\bgroup\let\\\endUnderline}
\def\endUnderline{\vphantom{y}\egroup\smash{\underline{\box0}}\\}
\def\|{\verb|}

\received{2015}{5}{18}
\accepted{2015}{11}{8}

\usepackage[varg]{txfonts}%%!!
\makeatletter%
\input{ot1txtt.fd}
\makeatother%

\usepackage{preprint}
\PROheadtitle{To appear in Journal of Information Processing, 24(2), 2016}

\begin{document}

\title{A Declarative Extension of Parsing Expression Grammars for Recognizing Most Programming Languages}

\affiliate{Z}{Zuken Inc. He worked this paper when he studied in Yokohama National University}

\affiliate{YNU}{Graduate School of Electronic and Computer Engineering,
Yokohama National University,
79-1 Tokiwadai, Hodogaya-ku, Yokohama 240--8501 JAPAN}

\author{Tetsuro Matsumura}{Z}
\author{Kimio Kuramitsu}{YNU}[kimio@ynu.ac.jp]

\begin{abstract}
Parsing Expression Grammars are a popular foundation for describing syntax. Unfortunately, several syntax of programming languages are still hard to recognize with pure PEGs. Notorious cases appears: typedef-defined names in C/C++, indentation-based code layout in Python, and HERE document in many scripting languages. To recognize such PEG-hard syntax, we have addressed a declarative extension to PEGs. The "declarative" extension means no programmed semantic actions, which are traditionally used to realize the extended parsing behavior. Nez is our extended PEG language, including symbol tables and conditional parsing. This paper demonstrates that the use of Nez Extensions can realize many practical programming languages, such as  C, C\#, Ruby, and Python, which involve PEG-hard syntax.
\end{abstract}


\maketitle

%1

\section{Introduction}

Parsing Expression Grammars\cite{POPL04_PEG}, or PEGs, are a popular foundation for describing programming language syntax\cite{PLDI06_Rats,PPPJ09_Fortress}. Indeed, the formalism of PEGs has many desirable properties, including deterministic behaviors, unlimited look-aheads, and integrated lexical analysis known as scanner-less parsing. Due to these properties, PEGs allow grammar developers to avoid the {\em dangling if-else} problem and to express the {\em contextual tokens} and the {\em nested block comment}, which are known problems in traditional LR($k$) and LL($k$) grammars.

Despite the powerful features of PEGs, several {\em real} syntax used in popular programming languages is hard to express. This problem comes chiefly from context-sensitive syntax, whose meanings are changed depending on the parsing context. Typical examples of such syntax are:

\begin{itemize}
\item Typedef-defined name in C/C++ \cite{POPL04_PEG,PLDI06_Rats}
\item HERE document appearing in Perl, Ruby, and other many scripting languages
\item Indentation-based code layout appearing in Python and Haskell\cite{POPL13_Indentation}
\item Contextual keywords used in C\# and other evolving languages\cite{OOPSLA06_AspectJ}
\end{itemize}

Technically, a language grammar involving such context-sensitive syntax is implemented with semantic actions\cite{Yacc}, an embedded code that is hocked to execute the extended action at the parsing time. Since semantic actions are written in a host language of the generated parser, the use of semantic actions may invalidate the declarative property of PEGs, thus resulting in reduced reusability of grammars. As a result, many developers need to redevelop grammars for their software engineering tools. 

In this paper, we propose a declarative extension of PEGs for recognizing context-sensitive syntax. The "declarative" extension means no arbitrary semantic actions that are written in a general purpose programming language. In our proposal, a variety of semantic actions are abstracted to two modeled states:

\begin{itemize}
\item {\em Symbol Table} -- a table that manages a list of symbols that are differently treaded in a specific context
\item {\em Parsing Condition} -- a Boolean flag that dynamically switches the parsing behaviors. 
\end{itemize}

Nez is an extended PEG language that is designed to provide additional operators to handle the above states in parsing contexts of PEGs. In this paper, we call {\em Nez extensions} to distinguish from {\em pure} PEGs operators.

Using Nez, we have performed extensive case studies by specifying various popular programming languages. The grammars that we have developed for now include C, Java, C\#, JavaScript, Lua, Ruby, and Python. Since these grammars include many context-sensitive syntax, we conclude that Nez extensions provide improved expressiveness for programming languages in practical manners. 

The remainder of this paper proceeds as follows. Section 2 describes PEG-hard syntax patterns. Section 3 presents a language design for Nez. Section 4 demonstrates case studies with our emphasis on Nez extensions. Section 5 briefly reviews related work. Section 6 concludes the paper. The tools and grammars that are presented in this paper are available online at \url{http://nez-peg.github.io/}

\section{Background and Problem Statement}

\subsection{PEGs}

PEGs are a recognition-based foundation for describing syntax, formalized by Ford\cite{POPL04_PEG}. While PEGs are relatively new, most of their notations are familiar, coming from EBNF (e.g., productions and recursive nonterminals) and regular expressions ({\em kleene} operators such as {\tt ?} and {\tt *}). Figure \ref{fig:math} shows a PEG, which expresses the basic mathematical notations.

\begin{figure}[tb]
\begin{framed}
\begin{tabular}{p{1.0cm} l}
{\tt Expr} & \verb|= Sum| \\
{\tt Sum} & \verb|= Product (( '+'  / '-' ) Product )*| \\
{\tt Product} & \verb|= Value (( '*' / '/' ) Value )*| \\
{\tt Value} & \verb|= [0-9]+ / '(' Expr ')' | \\
\end{tabular}
\end{framed}
\caption{Mathematical Operators in a PEG}
\label{fig:math}
\end{figure}

The interpretation of PEGs significantly differs from CFGs in that PEG's choice is ordered. That is, the first subexpression of a choice is always matched first, and the next subexpression is attempted only if the first fails. This brings us to deterministic parsing behavior, which is regarded as desirable for parsing non-natural languages. The ordered choice, on the contrary, disallows the {\em left recursion}, because the deterministic interpretation of the left recursion results in unlimited looping. The PEG example in Figure \ref{fig:math} is defined in a form of eliminating left recursions. 

The expressiveness of PEGs is almost similar to that of {\em deterministic} CFGs (such as LALR and LL family) . In general, PEGs can express all LR grammar languages, which are widely used in a standard parser generator such as Lex/Yacc.  In addition, PEG's syntactic predicates (\& and !) provide us with the expressiveness of unlimited look-aheads, suggesting that PEGs are more expressive than CFGs. As a result, PEGs can recognize non context free languages such as  $\{a_n$ $b_n$ $c_n$ $|$ $n$ $>$ $0\}$. 

\subsection{Context-Sensitive Syntax}

PEGs are very powerful, but, in practice, are not able to express all programming language syntax. This mostly comes from context-sensitive syntax, where the meaning of symbols may vary in different contexts. Note that the same limitation commonly exists in CFG-based grammars. This subsection demonstrates typical examples of context-sensitive syntax. 
% Note that there may be other syntax patterns that appear in a single unique language. We would not like to argue a single appearance pattern for generality.

\subsubsection{Typedef-defined name (in C/C++)}

Typedef-defined name in C/C++ is a typical example of context-sensitive syntax. The identifier is simply supposed as a sequence of word characters; for example, {\tt T} is a legal identifier in C/C++. On the other hand, the following {\tt typedef} statement allows the users to declare {\tt T} as a new type name. 

{\small \begin{framed} \begin{verbatim}
  typedef unsigned int T;
\end{verbatim} \end{framed}}

Once {\tt T} is a declared type, {\tt T} is not regarded as an identifier. In general, the interpretation of an identifier class should be performed in the phase of semantic analysis, not in syntactic analysis. However, in C/C++, we need to produce a different syntax tree depending on the contextually varying type of {\tt T}. That is, an expression {\tt (T)-1}  can be parsed differently:

\begin{itemize}
\item a subtract operator of a variable T and the number 1, or
\item a type cast of the number 1 to the type {\tt T}
\end{itemize}

Note that Java avoids this problem by a careful design of language syntax. The casting {\tt (T)-1} is disallowed by restricting that the unary + and - is only available for primitive number types. 

\subsubsection{HERE document (in Ruby, Bash, etc.)}

The {\em HERE document} is a string literal for multiple lines, widely adopted in scripting languages, such as Bash, Perl, and Ruby. While there are many variations, the users are in common allowed to define a delimiting identifier that stands for an end of lines. In the following, the symbol {\tt END} is an example of the user-defined delimiting identifier. 

{\small \begin{framed} \begin{verbatim}
print <<END      
  the string
  next line
END
\end{verbatim} \end{framed}}

In PEGs, the syntax of a HERE document can be specified with multi-lines that continue until the nonterminal {\tt DELIM} is matched in the head of a line. However, we cannot assume specific keywords or possible identifiers to define {\tt DELIM}, since the user-defined delimiting identifiers are totally unpredictable.

\subsubsection{Indentation-based layout (in Python, Haskell, etc.)}

Indentation-based code layout is a popular syntax that uses the depth of indentations for representing the beginning and the end of a code block. Typically, Python and Haskell are well known for such indentation-based code layout, a large number of other languages including YAML, F\#, and Markdown also use indentation.

{\small \begin{framed} \begin{verbatim}
mapAccumR f = loop
     where loop acc (x:xs) = (acc’’, x’ : xs’)
             where (acc’’, x’) = f acc’ x
                   (acc’, xs’) = loop acc xs
           loop acc [] = (acc, [])
\end{verbatim} \end{framed}}

The problem with PEGs is that the user is allowed to use arbitrary indentation for their code layout. As with in delimiting identifiers in the HERE documents, it is hard to prepare all possible indentations for code layout. 

\subsubsection{Contextual keyword in C\#}

Popular programming languages are long standing, and then evolve to meet the user's demands. In the language evolution, adding a new keyword is often considered to identify a new syntactical structure. On the other hand, a backward compatibility problem inevitably arises since the added keyword might have already been used as identifiers in legacy code.

A {\em contextual keyword} is used to avoid the compatibility problem. For example, C\#5.0 newly added the {\tt await} keyword, only available in the {\tt async} qualified method. 

{\small \begin{framed} \begin{verbatim}  
  async Task<string> GetPageAsync(string path)
  { 
    HttpClient c = new HttpClient();
    return await c.GetStringAsync(uri + path);
  }
\end{verbatim} \end{framed}}

As with in typedef names, the different meaning of {\tt await} needs to produce different syntax trees in the phase of syntactic analysis. It is also hard to specify different syntax analysis depending on given contexts. 

\subsection{Semantic Actions}

Semantic actions are a programmed code embedded in grammars definition, which is hocked to perform extra processing in parsing contexts. The following is a Rats$!$'s grammar fragment that shows an example of semantic actions \verb|&{ ... }|\cite{PLDI06_Rats}. The parse result of the nonterminal {\tt Identifier} is assigned to a variable {\tt id} in the semantic action and checked by the method {\tt isType()} of a global state {\tt yyState}. 

{\small \begin{framed} \begin{verbatim}
 TypedefName = id:Identifier &{
      yyState.isType(toText(id))
 }
\end{verbatim} \end{framed}}

%\end(lstlisting}

As shown above, semantic actions can use a general-purpose programming language, leading to richer operations including AST construction. As a result, semantic actions are most commonly used in today's parser generators to extend the expressiveness of grammars. 

An obvious problem with semantic actions is that the grammar definition depends tightly on a parser implementation language, and lacks the declarative properties of grammar definitions. This is really unfortunate, because well-defined grammars are potentially available across many parser applications such as IDEs and other software engineering tools. 

\section{Language Design of Nez}

Nez is a PEG-based grammar specification language that provides pure and declarative notations for AST constructions and enhanced matching for context-sensitive syntax. In this section, we focus on Nez extensions\footnote{We have planned to revise the Nez operators to improve the usability.  The {\tt def} operator was obsolete and replaced with the similar {\tt symbol} operator. The replaced operator ensures the same expressiveness with the {\tt def} operator. } for enhanced matching of context-sensitive syntax.

\subsection{Overview}

Nez is a PEG-based language that provides declarative notations for describing syntax without {\em ad hoc} semantic actions. The extensions range from AST constructions to enhanced matching. 
Figure \ref{fig:nez} shows an abstract syntax of Nez language. 

Nez operators are categorized as follows:

\begin{itemize}
\item {\em PEG Operators} -- matching operators based on PEGs
\item {\em AST Constructions} -- manipulating abstract syntax tree representations with parsed results 
\item {\em Symbol Tables Handlers} -- handling the global state, called symbol tables, in parsing contexts.
\item {\em Parsing Conditions} -- switching parser behavior depending on a given conditions. 
\end{itemize}

We have designed the capability of AST constructions in a decoupled way from any matching capability. That is, they are orthogonal to each other; any constructed ASTs do not influence matching results, while any matching results can be incorporated to tree manipulations. In this paper, we highlight the extended matching capability with symbol table handlers and parsing conditions, due to the space constraints. Further information on the AST construction can be referred to our report\cite{ASTMachine}.

\begin{figure}[tb]
\begin{center}
\begin{tabular}{lrll} %\hline
$e$ &  \verb|      ::= | & $\epsilon$ & { : empty} \\ 
& \verb#|  # & $A$ & { : non-terminal } \\
& \verb#|  # & $a$ & { : terminal character} \\
%& \verb#|  # & \verb|.| & { : any character } \\
& \verb#|  # & $e ~ e' $ & { : sequence| } \\
& \verb#|  # & $e ~ / ~ e' $ & { : prioritized choice } \\
& \verb#|  # & $e?$ & { : option } \\
& \verb#|  # & $e*$ & { : repetition } \\
%& \verb#|  # & $e+$ & { : (one more) repetition } \\
& \verb#|  # & \verb|&|$e$ & { : and predicate } \\
& \verb#|  # & \verb|!|$e$ & { : not predicate } \\
& \verb#|  # & $\{ e \}$ & { : AST constructor } \\
& \verb#|  # & $\$(e)$ & { : AST connector } \\
& \verb#|  # & \verb|#x| & { : AST tagging } \\
% & \verb#|  # & \verb|`x`| & {\tt : string-replace } \\
& \verb#|  # & \verb|<def| $~T~e$ \verb|>| & { : symbol definition} \\
& \verb#|  # & \verb|<exists| $~T$ \verb|>| & { : symbol existence  } \\
& \verb#|  # & \verb|<match| $~T$ \verb|>| & { : symbol match } \\
& \verb#|  # & \verb|<is| $~T$ \verb|>| & { : symbol equivalence } \\
& \verb#|  # & \verb|<isa| $~T$ \verb|>| & { : symbol containment  } \\
& \verb#|  # & \verb|<block| $~T~e$ \verb|>| & { : nested table scoping } \\
& \verb#|  # & \verb|<local| $~T~e$ \verb|>| & { : isolated table scoping } \\
& \verb#|  # & \verb|<if| $~C$ \verb|>| & { : condition testing} \\
& \verb#|  # & \verb|<on| $C~e$ \verb|>| & { : evaluation on the condition C} \\
% & \verb#|  # & \verb|<off | $C$ \verb|>| & {\tt : turn C off } \\
\hline
\end{tabular}
\end{center}
\caption{An abstract syntax of Nez language}
\label{fig:nez}
\end{figure}

\subsection{Symbol Tables}

Symbol table is a global state used to maintain strings, whose meaning is specialized in parsing contexts. We call such strings {\em symbols} in this paper. Nez supports multiple symbol tables in a grammar. Let $T$ be a table identifier to distinguish a symbol table from others. 

Nez newly defines the following operations on the symbol table $T$:

\begin{itemize}
\item $\verb|<def| ~T~e\verb|>|$ -- symbol definition by extracting a string matched by the subexpression $e$, and then store it as a symbol to the table $T$
\item $\verb|<match| ~T\verb|>|$ -- match the latest-defined symbols in the table $T$
\item $\verb|<is| ~T\verb|>|$ -- equals the latest-defined symbol in the table $T$
\item $\verb|<isa| ~T\verb|>|$ -- contains one of stored symbols in the table $T$
\item $\verb|<exists| ~T\verb|>|$ -- testing the existence of stored symbols in the $T$ table
\item $\verb|<local| ~T~e\verb|>|$ -- isolated local scope of $T$ for the subexpression $e$
\item $\verb|<block| ~T~e\verb|>|$  -- nested local scope of $T$ for the subexpression $e$
\end{itemize}

Let us show how a symbol table works with parsing expressions. To start, we consider the absence of any symbol tables. The production XML is intended to accept \verb|<tag> .. </tag>|.

{\small \begin{verbatim}

 XML  = '<' NAME '>' XML? '</' NAME '>'
 NAME = [A-z] [A-z0-9]*

\end{verbatim}}

An obvious problem is that the parsed closing tag can be different from the parsed opening tag because the production \nt{NAME} matches arbitrary names. That is, the production \nt{XML} accepts an illegal input such like \verb|<a></b>|. Since a set of possible names are infinite, it is hard for pure PEGs to ensure the equivalence of opening tags and closing tags. 

Now we consider the introduction of a symbol table, named TAG, to maintain names parsed at the opening tag. 

{\small \begin{verbatim}

 XML  = '<' <def TAG NAME> '>' XML? '</' <is TAG> '>'
 NAME = [A-z] [A-z0-9]*

\end{verbatim}}

Due to the stored opening tag, we can check the equivalence of the closing tag. However, \verb|<is TAG>| only accepts the latest-defined symbol in {\tt TAG}. That is, nested tags, such as \verb|<a><b></b></a>|, are still unacceptable. If we replace \verb|<is TAG>| with \verb|<isa TAG>|, then we can match either 'a' or 'b' but the proper order is not guaranteed. (That is, \verb|<isa TAG>| is a bad example.)

The constructor $\verb|<block TAG |~e\verb|>|$ is introduced to declare a nested local scope of symbol tables. The scope means that all symbols that defined in the subexpression $e$ are only available in the context of evaluating $e$.  In other words, any symbols defined in $e$ are not available outside the {\tt block} constructor. The following is a scoped-modification to match nested tags correctly.

{\small \begin{verbatim}

 INNER = <block TAG XML>
 XML  = '<' <def TAG NAME> '>' INNER? '</' <is TAG> '>'
 NAME = [A-z] [A-z0-9]*

\end{verbatim}}

Nez provides another scoping notation, $\verb|<local| ~T~e\verb|>|$. The difference is that the {\tt local} creates an isolated scope for a specific table $T$. The isolated scoping means that symbols defined outside are not referred to from the inside scope. In the above, the replacement of the block scope with $\verb|<local TAG XML>|$ produce the same matching result. 

\subsection{Parsing Condition}

The idea of parsing condition comes from both {\em conditional compilation} and {\em semantic predicates}. The parsing conditions are similar to directives in conditional compilation and switch the parsing behavior by whether the parsing flag is true or not. 

Nez supports multiple parsing conditions, identified by the user-specified flag names.  Let $C$ be a parsing flag name. The notation $\verb|<if| ~ C \verb|>|$ is a determination of the parser behavior depending on whether $C$ is true or not. That is, an parsing expression $\verb|<if| ~C\verb|>| ~ e$ means that the expression $e$ is attempted only if $C$ is {\em true}.  The syntactic predicate $!$ is allowed for a parsing flag, which is simply a negation of $C$. That is, $\verb|<if| ~!C\verb|>| ~ e$ means that the expression $e$ is attempted only if the parsing flag $C$ is {\em false}. The expressions $e_1$ and $e_2$ can be distinctly switched with a choice by $\verb|<if|~C\verb|>| ~ e_1 ~/~ \verb|<if|~!C\verb|>| ~ e2$. 

Here is an example of a parsing condition {\tt NL} that switches the inclusion of new lines in white spaces. The production {\tt WS} is regarded as \verb|[ \t\n]| if {\tt NL} is true, and \verb|[ \t]| if not. 

{\small \begin{verbatim}

  WS  = [ \t] / <if NL> [\n]

\end{verbatim}}

The parsing conditions are not static, rather we switch them on/off in parsing contexts. Switching conditions is only controlled by the following two constructors:

\begin{itemize}
\item $\verb|<on|~C~e\verb|>|$ -- the expression $e$ is evaluated under the condition that C is true
\item $\verb|<on|~!C~e\verb|>|$ -- the expression $e$ is evaluated under the condition that C is false
\end{itemize}

Note that $\verb|<on| ~C~e\verb|>|$ and $\verb|<on| ~!C~e\verb|>|$ are not an action to be performed, but a condition declaration to be satisfied for the subexpression $e$. We allow nested declarations. If undeclared, conditions are regarded as true by default.

{\small \begin{verbatim}

  <on NL WS> // on
  <on !NL WS> // off
  <on NL <on !NL WS>>  // nested

\end{verbatim}}

Note that the parsing condition is a global state and sharable across productions. In the above example, the condition {\tt NL} used in the production {\tt WS} is switched from arbitrary contexts of $\verb|<on NL|~e~\verb|>|$ whose $e$ involves the nonterminal {\tt WS}.  

\subsection{Semantics}

The semantics of the Nez extensions is built on the formal semantic of PEGs, presented in \cite{POPL04_PEG}. Formally, a Parsing Expression Grammar, $G$, is a 4-tuple $G = (V_N, V_T, R, e_s)$, where $V_N$ is a finite set of nonterminal, $V_T$ is a finite set of terminal, $R$ is a finite set of rules, $e_s$ is a start expression. Each rule, $A = e$, is a mapping from nonterminal $A$ to parsing expression $e$. This mapping is written as $R(A)$. 

Let $x, y, z, w$ be a string. The symbol table $T$ is defined as a recursive ordered pair of string. The qualifier “recursive” means that a list $(x, y, z)$ equals to a nested pair $(x, (y, z))$. $()$ stands for an empty list. The pair $(w, T)$ stands for a new pair where a string $w$ is added to $T$. The function $top(T)$ is defined as $top(T) = x$ where $T = (x, y, z, ... )$.

Let $xy$ be a concatenation of $x$ and $y$. We assume the single table case without the loss of generality. The semantics of PEGs with $T$ is given as a relation $(e, xy, T) \Rightarrow (n, y, T' )$, where $e$ is a parsing expression, $xy$ is an input string, $n$ is a step counter, and $y$ is an unconsumed string. That is, the expression $e$ consumes the string $x$. If the expression fails, we write $\bullet$ for the remaining string. The $T$ and $T'$  represent a certain status of the symbol table, which may be different. 

The relation $\Rightarrow$ is defined inductively as shown in Figure \ref{fig:semantics}. As in \cite{POPL04_PEG}, we use the abstract form of PEG syntax by omitting syntax sugars. In addition, we regard the condition name $C$ as a string that can be stored in the table $T$. 

\begin{figure}[tb]
\begin{eqnarray*}
\textrm{Empty} & : & (\epsilon, x, T) \Rightarrow (1, x, T) \\
\textrm{Terminal} & : & (a, ax, T) \Rightarrow (1, x, T) \\
% Terminal(2) & : & (a, bx, T) \Rightarrow (1, \bullet, T) ~ \textbf{if} ~ a \ne b\\
\textrm{NonTerminal} & : & (A, xy, T) \Rightarrow (n+1, y, T') ~~ \textbf{if} ~~ (R(A), xy, T) \Rightarrow (n, y, T') \\
\textrm{Sequence} & : & (e_1~e_2, xyz, T) \Rightarrow (n_1+n_2+1, z, T'') \\
& & \textbf{if} ~~ (e_1, xy, T) \Rightarrow (n_1, y, T') ~~ \textbf{and} ~~ (e_2, yz, T') \Rightarrow (n_2, z, T'') \\
\textrm{Choice} & : & (e_1/e_2, xy, T) \Rightarrow (n_1+1, y, T') \\
& & \textbf{if} ~~ (e_1, xy, T) \Rightarrow (n_1, y, T') \\
\textrm{Choice(2)} & : & (e_1/e_2, xy, T) \Rightarrow (n_2+1, y, T') \\
& & \textbf{if} ~  (e_1, xy, T) \Rightarrow (n_1, \bullet, T') ~~ \textbf{and} ~~  (e_2, xy, T) \Rightarrow (n_2, y, T') \\
\textrm{Repetition} & : & (e*, xyz, T) \Rightarrow (n_1+n_2+1, y, T'') \\
& & \textbf{if} ~  (e_1, xyz, T) \Rightarrow  (n+1, yz, T') ~~ \textbf{and} ~~  (e*, yz, T') \Rightarrow (n_2, z, T'') \\
\textrm{Not} & : & (!e, x, T) \Rightarrow (n+1, x, T') ~~ \textbf{if} ~  (e, x, T) \Rightarrow (n, \bullet, T') \\
\textrm{Def} & : & (\langle def ~T ~e\rangle, xy, T) \Rightarrow (n+1, y, (x, T)) \\
& & \textbf{if} ~  (e, xy, T) \Rightarrow (n, y, T)  \\
\textrm{Block} & : & (\langle block ~T~e\rangle, xy, T) \Rightarrow (n+1, y, T) \\
& & \textbf{if} ~  (e, xy, T) \Rightarrow (n, y, T')  \\
\textrm{Local} & : & (\langle local ~T~e\rangle, xy, T) \Rightarrow (n+1, y, T) \\
& & \textbf{if} ~  (e, xy, ()) \Rightarrow (n, y, T')  \\
\textrm{Exists} & : & (\langle exists ~T \rangle, x, T) \Rightarrow (n+1, x, T) 
~~ \textbf{if}  ~~  \exists~w  ~~\textbf{such that}~~ w \in T \\
\textrm{Match} & : & (\langle match ~T \rangle, xy, T) \Rightarrow (n+1, y, T) \\
& & \textbf{if} ~  (x, xy, T) \Rightarrow (n, y, T) ~~ \textbf{and} ~~  x = top(T) \\
\textrm{Is} & : & (\langle is ~T \rangle, xy, T) \Rightarrow (n+1, y, T) \\
& & \textbf{if} ~  (e^T, xy, T) \Rightarrow (n, y, T) ~~ \textbf{and} ~~  x = top(T) \\
\textrm{Isa} & : & (\langle isa ~T \rangle, xy, T) \Rightarrow (n+1, y, T) \\
& & \textbf{if} ~  (e^T, xy, T) \Rightarrow (n, y, T) ~~ \textbf{and} ~~  x \in T \\
\textrm{If} & : & (\langle if ~C \rangle, x, T) \Rightarrow (n+1, x, T)  ~~ \textbf{and} ~~    C \in T \\
\textrm{On} & : & (\langle on ~C~e \rangle, xy, T) \Rightarrow (n+1, y, T) \\
& & \textbf{if} ~  (e, xy, (C, T)) \Rightarrow (n, y, T')  \\
\end{eqnarray*}
\caption{Semantics}
\label{fig:semantics}
\end{figure}

In Nez, we presume that strings in the table $T$ are all extracted from the same expression. That is, two different table definitions ($\verb|<def|~T~e\verb|>|$ and $\verb|<def|~T~e'\verb|>|$ such that $e \ne e'$) are not allowed. We write $e^T$ for representing an expression that is defined in $\verb|<def|~T~e\verb|>|$. More importantly, symbol matching \verb|<is|$~T$\verb|>| and symbol containment \verb|<isa|$~T$\verb|>| are based on the sub-string extraction by $e^T$ in order to avoid unintended substring matching. 

\section{Case Studies and Experiences}

We have developed many grammars ranging from programming languages to data formats. All developed grammars are available online at \url{http://nez-peg.github.io/}. This section reports our experiences throughout grammar developments with Nez. 

\subsection{Summary}

To evaluate the language design of Nez, we have performed extensive case studies by developing grammars for major programming languages. Our examined grammars are developed by the following approaches:

\begin{itemize}

\item C -- Based on two PEG grammars written in Mouse and Rats$!$. Semantic actions for handling the {\tt typedef} statement are ported into Nez's symbol table. 

\item Java8 -- Ported from Java8 grammar written in ANTLR4\footnote{https://github.com/antlr/grammars-v4/blob/master/java8/Java8.g4}

\item JavaScript -- Based on JavaScript grammar for PEG.js\footnote{https://github.com/pegjs/pegjs/blob/master/examples/javascript.pegjs}. 

\item C\# -- Developed from the scratch, referencing C\#5.0 Language Specification (written in a natural language)

\item Lua -- Developed from the scratch, referencing Lua 5.1 Reference Manual. 

\item Ruby -- Developed from the scratch and in part referred as the yacc grammar for CRuby.

\item Parser -- Ported from Python 2.7 abstract grammar\footnote{https://docs.python.org/2.7/library/ast.html}

\item Konoha --Developed from the scratch, based on Konoha paper\cite{Konoha}

\item MinCaml --Developed from the scratch, based on MinCaml paper\cite{FDPE05_MinCaml}

\item Haskell -- Developed from the scratch. 

\end{itemize}

Table \ref{table:grammars} shows a list of developed grammars.The column labeled "\#" indicates the number of production rules, implying the complexity of grammars. 
Table \ref{table:grammars} confirms a substantial trend in the expressiveness of PEGs, while our developed grammars are in part incomplete. Only two languages (including Java8 and MinCaml) can be fully specified with {\em pure} PEGs. This suggests that most of programming languages require semantic actions or equivalent extensions such as the Nez extensions in order to parse with PEGs.

%The column "Pass" indicates the quality of developed grammars, by measuring a pass ratio of software testing for files randomly collected from the Web. Untested grammars remain blank. The column labeled "Nez ext." indicates the types of Nez notations involved in grammars.

\begin{table}[tb]
\caption{List of developed grammars for programming languages}
\label{table:grammars}
\begin{center}
\begin{tabular}{lrl}
Language & \# of Rules & Nez ext.\\ \hline
%% Bash & 109 &  & \verb|<|def\verb|>| \verb|<|is\verb|>| \verb|<|block\verb|>|  \\
C & 101 &  \verb|<|def\verb|>| \verb|<|isa\verb|>| \\
C\#5.0 & 454  & \verb|<if>| \verb|<on>| \\
Haskell98 & 110  & \verb|<|def\verb|>| \verb|<|is\verb|>| \verb|<|block\verb|>| \verb|<|if\verb|>| \\
Java8 & 160  &  \\
JavaScript & 132  & \verb|<if>| \verb|<on>| \\
Konoha & 124 & \verb|<if>| \verb|<on>| \\
MinCaml & 65  & \\
Lua & 96  & \verb|<|def\verb|>| \verb|<|is\verb|>| \verb|<|block\verb|>| \verb|<|if\verb|>| \\
Python & 55  & \verb|<|def\verb|>| \verb|<|match\verb|>| \verb|<|block\verb|>| \verb|<|if\verb|>| \\
Ruby & 200  & \verb|<|def\verb|>| \verb|<|is\verb|>| \verb|<|block\verb|>| \verb|<|on\verb|>| \verb|<|if\verb|>| \\
\hline
\end{tabular}
\end{center}
\end{table}

Throughout our case studies on Nez extensions, we have obtained positive forecasts for expressing each of full language specifications. A significant exception is Haskell. As described below, Haskell's code layout is too amended to programmers. In addition, Haskell supports syntax extensions, which allow the users to even change the precedence of binary operators. This extensibility is not good for PEG's deterministic parsing based on the operator associativity.  

The remainder of this section describes each language syntax that focuses on the Nez extensions. Note that all examples presented in this subsection are modified for improved readability. 

\subsection{C and Typedef-Defined Name}

Parsing {\tt typedef}-defined names is a classic problem of parser generators even with LALR and LL families. Figure \ref{fig:c} shows excerpted productions from our C grammar. The production \nt{TypeDef} describes a syntax of the {\tt typedef} statement with defining matched symbols in the TYPE table. The production \nt{TypeName} is defined to first match built-in type names and then match one of symbols stored in the TYPE table.

\begin{figure}[t]

{\small \begin{framed} \begin{verbatim}

 W  = [A-ZA-z_0-9]
 S  = [ \t\r\n]
  
 TypeDef  
    = 'typedef' S* TypeName S* <def TYPE W+> S* ';'

 TypeName 
    = BuildInTypeName / <isa TYPE>

 BuiltInType
    = 'int' !W / 'long' !W / 'float' !W ...
    
\end{verbatim} \end{framed}}

\caption{Productions for {\tt typedef} and type names in C grammars}
\label{fig:c}

\end{figure}

If the scope of type-defined names were global only, the production \nt{TypeName} could work in any contexts. In reality, the {\tt typedef} statement allows nested local scoping in a mixed manner with local variables. For example, the following is an legal C code: 

{\small \begin{verbatim}

typedef int T;

/* T is a type name */

int main() {
  int T = 0;
  /* Here, T is a variable */
  {
    typedef double T;
    /* Here, T is a type name */
    printf("T=%f\n", (T)-1);
  }
  /* Again, T is a variable */
  printf("T=%d\n", (T)-1);
  return T; 
}
/* Again, T is a type name */

\end{verbatim}}

To parse the above concisely, we need to maintain all local variable names with another symbol table, as well as to introduce the nested scoping with \verb|<block>|. Let VAR be a symbol table for local variables. The following is a modified \nt{TypeName} production to test the local variable name. 

{\small \begin{verbatim}

 TypeName 
    = BuildInTypeName / !<isa VAR> <isa TYPE>

\end{verbatim}}

This modification however still has difficulty in handling the inner nested {\tt typedef} statement. Since Nez provides no supports for distinguishing table types, we cannot perform a concise parsing of duplicated names. However, the {\tt typedef} statement is mostly used at the top level of source code. This is why we consider it not to be serious in the most practical cases.

\subsection{HERE Document}

The HERE document is a popular syntax of multiple line strings, by allowing the user to define a delimiting identifier of the end of those lines. The user-defined identifier, as with in typedef-defined names, can be maintained in a symbol table. Figure \ref{fig:rb} shows a fragmentation of the Ruby grammar.

The statement definitions involving \nt{HereDocu} are defined inside a local scope. The production \nt{HereDocu} matches a delimiting identifier to store in the DELIM table. The body of HERE document follows the statement, depending on the DELIM table. We use \verb|<exists DELIM>| to test the existence of table entries. If it exists, we parse subsequent lines until the head of the line starts with \verb|<is DELIM>|. 

Note that the defined delimiting identifier is available only inside \nt{StatementDocu} and \nt{Document} productions. If a statement includes another statement, the isolation of scope is necessary. In such cases, we may use \verb|<local>|, instead of  \verb|<block>|.

\begin{figure}[tb]

{\small \begin{framed} \begin{verbatim}

 NL = '\r\n' / '\n'
 
 Statement
   = <block StatementDocu NL
            Document?   >

 HereDocu
   = '<<' <def DELIM W+ > 
     
 Document
   = <exists DELIM> 
    !<is DELIM> (!NL .)* NL <is DELIM> NL

\end{verbatim} \end{framed} }
\caption{Productions for HERE Document in Ruby}
\label{fig:rb}

\end{figure}

Ruby and many other scripting languages allows multiple HERE documents in a single statement. That is, the following is a legal code:

{\small \begin{verbatim}
 
 puts <<FIRST, <<SECOND
 ...
 FIRST
 ...
 SECOND
 
\end{verbatim}}

Nez provides no meta-variables for tables. Accordingly, we need to define a fixed number of tables in preparation. How many tables we need upfront depends on the language specification, while Ruby's reference manual does not mention the maximum number.  

{\small \begin{verbatim}

 HereDocu
   = '<<' (!<exists DELIM> <def DELIM W+ > )
       /  (!<exists DELIM2> <def DELIM2 W+ > )
       / ..
       
 Document
   = <exists DELIM> 
    !<is DELIM> (!NL .)* NL <is DELIM> NL
     ( <exists DELIM2>
      !<is DELIM2> (!NL .)* NL <is DELIM2> NL
      ...
     )?

\end{verbatim}}

\subsection{Contextual Keywords}

A contextual keyword is used to avoid the backward compatibility problem with language evolutions. An added new keyword is used to provide a specific meaning in a specific context of code; outside the context, programmers can still use it as an identifier. 
Figure \ref{fig:ck} shows a list of contextual keywords in C\#5.0, implying that the evolution of C\# relies largely on many contextual keywords. The same ideas are extensively discussed in many cases of language evolutions, including the future version of JavaScript and  C++0x.  

\begin{figure}[tb]

\begin{framed}
\noindent {\tt 
add
alias
ascending
async
await
descending
dynamic
from
get
global
group
into
join
let
orderby
partial
remove
select
set
value
var
where
yield}
\end{framed}

\caption{Contextual Keywords in C\#5.0}
\label{fig:ck}

\end{figure}

Fundamentally, PEGs are based on scanner-less parsing, and allows any tokens to be recognized as either keywords or identifiers depending on its context. Let us recall the {\tt await} case, described in Section 2.2.4. The token {\tt await} can be regarded as either an identifier by \nt{Identifier} or a keyword by \nt{IdentiferNonAwait}: 
  
{\small \begin{verbatim}

 W = [A-z_0-9]
 /* await can be an identifier */
 Identifier = !Keyword [A-z_] W+
 Keyword    = 'abstract' !W / 'as' !W / 'base' !W ...

 /* await is a keyword */
 IdentifierNonAwait 
            = !KeywordAwait [A-z_] W+
 KeywordAwait 
            = 'abstract' !W / 'as' !W / 'await' !W 
             / 'base' !W ...  

\end{verbatim}}

The specification of contexts is the hard part. The keyword {\tt await} is only available in {\tt async} modified methods. The production \nt{MethodDecl} needs to dispatch two different cases depending on the {\tt async} modifier. 

{\small \begin{verbatim}

MethodDecl
    = 'async'Spacing MethodDeclAwaitContext
    / MethodDeclContext

\end{verbatim}}

The production \nt{MethodDeclContext} is a standard version of method declaration, including such language syntax as blocks, statements, expressions, and variables. A problem with specifying \nt{MethodDeclAwaitContext} is that we need to rewrite all await versions of these sub-productions that involving \nt{Identifier} and \nt{Keyword}. In our experience, the rewrites are needed for approximately 107 productions. As easily imagined, this approach would involve a considerable number of tedious specification tasks and would be then prone to errors.

Nez's conditional parsing, on the other hand, allows a single definition of the \nt{Keyword} production to be differently recognized on a giving condition. 

{\small \begin{verbatim}

 W = [A-z_0-9]
 
 Identifier = !Keyword [A-z_] W+
 Keyword    = 'abstract' !W / 'as' !W 
            / <if AWAIT> 'await' !W 
            / base !W ...  

\end{verbatim}}

In addition, a single \nt{MethodDeclContext} definition is also allowed. This reduces the duplication tasks of similar productions by hands. 

{\small \begin{verbatim}

MethodDecl
    = "async" __ <on AWAIT MethodDeclContext>
    / <on !AWAIT MethodDeclContext>

\end{verbatim}}

It is important to note that the conditional constructs such as $\verb|<if |C\verb|>|$ and $\verb|<on |C~e\verb|>|$ can be removed from grammars by converting into condition-specific nonterminals. Actually, Nez parser performs such conversions upfront. As a result, the conditional parsing is an extension for improving the productivity of specification tasks. 

\subsection{Indentation-based Code Layout}

Indentation-based code layout is a popular style of code layout, typically used in Python and Haskell. At the same time, it is a known fact\cite{POPL13_Indentation,Haskell14_Indentation} that CFGs and PEGs are not able to recognize it without semantic actions.

While the symbol table is not specifically designed to handle indentations,  we can store white spaces as a specialized symbol for representing indentation. To illustrate, let \nt{S} be a spacing production such that \verb|S = [ \n]|. 

An indentation of a line heading can be defined as a white spacing symbol on the INDENT table:

{\small \begin{verbatim}

    <def INDENT S*> Statement

\end{verbatim}}

Note that \verb|S*| is greedy matching that consumes all white spaces before \nt{Statement}.

Now we can test the same length of white spaces with \verb|<match INDENT>|, and a deeper indentation can be controlled by \verb|<match INDENT>| followed by one-and-more repetition of white spaces:

{\small \begin{verbatim}

    <match INDENT> S+ Statement

\end{verbatim}}

The Python-style nested indentation layout can be handled by nested scope of the INDENT table. Figure \ref{fig:py} is an excerpted grammar from our Python syntax, where the parsing condition {\tt LO} is used to switch either layout-sensitive or layout-insensitive styles of code.

\begin{figure}[tb]

{\small \begin{framed} \begin{verbatim}

/* LO: a flag for indent-based layout */

Layout 
    = <if LO> <def INDENT <match INDENT> S+>
    / S*

Block
    = EOS <on LO <block INDENT Statement*>>
    / <on !LO Statement>
    
Statement
    = IfStatement / WhileStatement / …

IfStatement
    = Layout 'if' Expression ':' Block
     (Layout 'else' ':' Block)?

WhileStatement
    = Layout 'while' Expression ':' Block

\end{verbatim} \end{framed}}
\caption{Fragment of Python Grammars}
\label{fig:py}
\end{figure}

In addition, Python's indentation has an offside rule, or an exception of indentation-based layout; the layout is ignored inside parenthesized expression.  For example, the indentation ahead of {\tt 2} is regarded as nothing: 

{\small \begin{verbatim}

if cond:
      a = (1 +
  2)

\end{verbatim}}

As the readers imagine, the offside rule is similar to contextual keywords, which implies that conditional parsing can switch behaviors. In this case, we express an offside rule by simply surrounding the parenthesized expression such as \verb|'(' <on !LO Expr> ')'| .

Haskell has a different style of indentation-based code layout, where the depth of indentation is determined by specific keywords such as {\tt let} and {\tt where}. That is, for example, the following indentation of \verb|y = b| must start with white spaces at the same position of \verb|x = a|. 

{\small \begin{verbatim}

  let x = a
       y = b

\end{verbatim}}

Apparently, we cannot extract white spaces for the INDENT table from a matched string. To express the Haskell-style indentation, we require a language-specific symbol handler, such as \verb|<defindent>|. While we experimentally support such language-specific handlers, we don't mention them due to the generality of the Nez extensions.

\section{Performance Study}

Linear time parsing is a central concern of backtracking parsers because backtracking may easily impose exponential costs in the worst cases. In PEGs, packrat parsing\cite{ICFP02_PackratParsing} is a known implementation method to avoid such potential exponential costs. However, the Nez extensions would give rise to a problem with packrat parsing, because the linear time guarantee of packrat parsing is based on the fact that PEG-based parsing is stateless. In other words, the trick of packrat parsing is the memoization of nonterminal calls at distinct positions to avoid redundant repeated calls. Apparently, the Nez extensions invalidates  the feature of stateless parsing, as we described in the previous sections. 

In terms of the lack of stateless features, a semantic action approach involves the same problem. In this light, Grimm, the author of Rats$!$, has assumed in \cite{PLDI06_Rats} that the state changes in parsing programming languages always flow forward the input and then previously memoized results need to be invalidated. That is, the linearity of packrat parsing can be preserved.  We examine this assumption with our developed grammars and Nez parser. 

\begin{figure}[tb]
\begin{center}
\includegraphics[width=7.5cm]{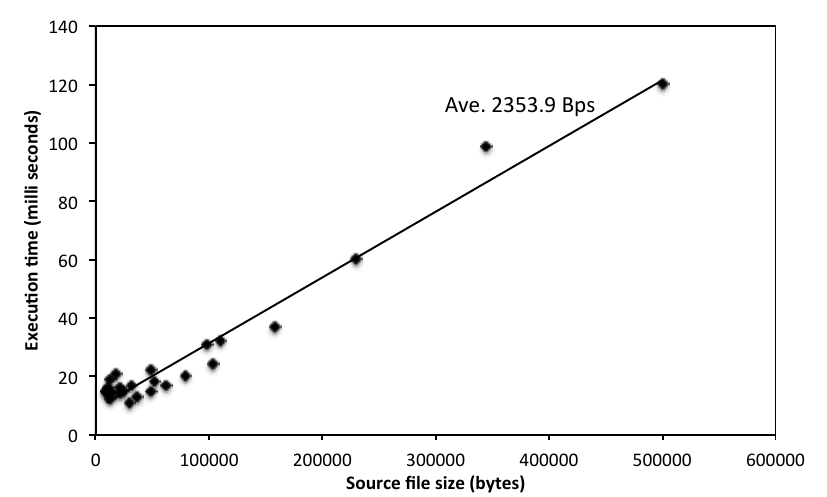}
\caption{Parsing Time in Java}
\label{fig:java}
\end{center}
\end{figure}

\begin{figure}[tb]
\begin{center}
\includegraphics[width=7.5cm]{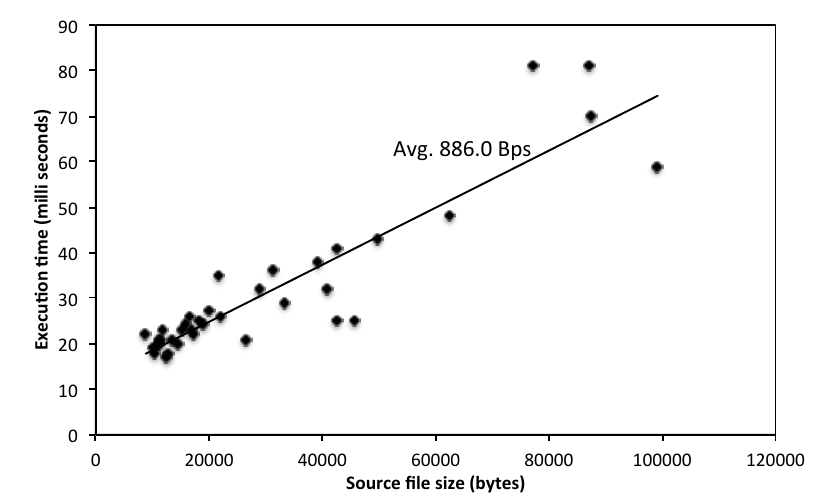}
\caption{Parsing Time in Ruby}
\label{fig:ruby}
\end{center}
\end{figure}

\begin{figure}[tb]
\begin{center}
\includegraphics[width=7.5cm]{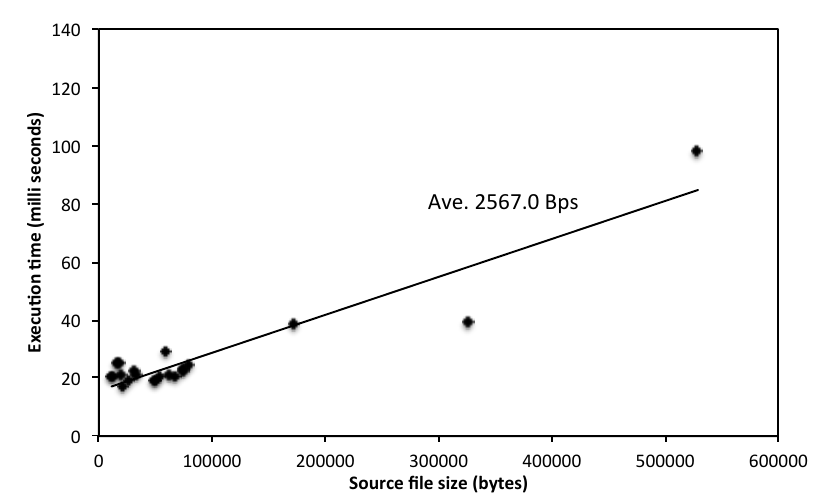}
\caption{Parsing Time in C\#}
\label{fig:csharp}
\end{center}
\end{figure}

Figures \ref{fig:java}, \ref{fig:ruby}, and \ref{fig:csharp} show the parsing time plotted against file sizes in, respectively, Java, Ruby, and C\#. These tests were measured on an Apple's Mac Book Air, with 2GHz Intel Core i7, 4MB of L3 Cache, 8GB of DDR3 RAM, running on Mac OS X 10.8.5 and Oracle Java Development Kit version 1.8. Tested files are collected from various open source repositories in order to examine different styles of coding. Tests run several times and we record the best time for each iteration. The execution time is measured by {\tt System.nanoTime()} in Java APIs. The Nez parser that we have tested is based on \cite{PRO101}. 

As explained int Section 4, the Java grammar contain no Nez extension and the C\# grammar contains parsing conditions, which are converted to condition free PEGs before parsing. These are stateless parsing, leading to no invalidation of packrat parsing. The Ruby grammar contains the symbol table handlers, \verb|<|def\verb|>| and \verb|<|is\verb|>|, which requires the treatment of state changes in packrat parsing. However, we haven't observed any significant difference on the linearity of parsing time, compared to other grammars. The reason is that the state changes operated by symbol definitions are perhaps localized and do not cause any significant invalidation of memoized results. We confirm Grimm's assumption with supporting evidence. 

%\begin{figure}[tb]
%\begin{center}
%\includegraphics[width=6.5cm]{mat-js}
%\caption{Parsing Time in JavaScript}
%\label{fig:js}
%\end{center}
%\end{figure}

\section{Related Work}

Due to the popularity of PEGs, many grammar developers have attempted the grammar specification for their interesting languages. While PEGs, in some sense, are  more powerful than CFGs, several limitations on their expressiveness have been pointed out in \cite{POPL04_PEG,PPPJ09_Fortress}. 

Since YACC\cite{Yacc} has been broadly accepted as a standard parser, the semantic action (embedded code in a grammar) is a traditional and common approach to enhance the expressiveness of formal grammars, such as LR($k$) and LL($k$)\cite{PLDI11_Antlr}. Grimm presents that the semantic actions can be applied even into the speculative parsing such as PEGs\cite{PLDI06_Rats}. More recently, most PEG-based parser generators (e.g., Mouse \cite{FI07_Mouse}, PEGTL\cite{PEGTL}, and PEGjs\cite{PEGjs}) have the semantic action supports for recognizing PEG-hard syntax, but the embedded action code depends on a host language of a parser. 

A few researchers have attempted to extend the expressive power of PEGs itself. Notably, Adams newly introduced Indent-Sensitive CFGs\cite{POPL13_Indentation} and its PEG-version\cite{Haskell14_Indentation} to recognize indentation-based code layout. The idea is based on constraint-based annotations on all nonterminals and terminals. As we described in Section 4.5, Nez can define an {\tt INDENT} table and provide similar (not the same) effects to the Indent-Sensitive CFGs, as shown in Table\ref{fig:indent}. 

\begin{table}[tb]
\begin{center}
\caption{Simple Correspondence between IS-CFGs and Nez INDENT table}
\label{fig:indent}
\begin{tabular}{ll} \hline
ISCFG & Nez/PEG \\ \hline
$e^{=}$ &  $\verb|<match INDENT>|~e$ \\
$e^{>}$ &  $\verb|<match INDENT>|~S$+ $~e$ \\
$e^{\ge}$ &  $\verb|<match INDENT>|~S$* $~e$ \\
$|e|$ &  $\verb|<block <def INDENT S*>| ~ e\verb|>|$ \\
$e^{*}$ &  $\verb|<local INDENT|~e\verb|>|$ \\ \hline
\end{tabular}
\end{center}
\end{table}
  
To our knowledge, Nez is the first attempt to the declarative supports for recognizing various context-sensitive syntax patterns, including limitations that Ford's first pointed out. 

\section{Conclusion}

Parsing Expression Grammars are a popular foundation for describing syntax. Unfortunately, pure PEGs find it difficult to recognize several syntax patterns appearing in major programming languages. Notorious cases include typedef-defined names in C/C++, indentation-based code block in Python, and HERE document used in many scripting languages. To recognize such PEG-hard patterns, we have designed Nez as a pure and declarative extension to PEGs. 

We present the language design of Nez, including symbol table handlers and conditional parsing. Using Nez, we have performed extensive case studies on programming language grammars, which include C, C\#, Java8, JavaScript, Lua, Python, Ruby, etc.  Our case studies indicate that the Nez extensions are a practical extension to improve the expressiveness of PEGs for recognizing major programming languages. Our developed artifacts will be available online, at \url{http://nez-peg.github.io/}. 
   
\begin{acknowledgment}
The authors thank the IPSJ PRO102 and PRO103 attendees for their feedback and discussions. 
\end{acknowledgment}

\bibliographystyle{ipsjsort-e}
\bibliography{../bib/parser,../bib/mypaper,../bib/url}  

\begin{biography}

\profile{Tetsuro Matsumura}{ received the master degree from Yokohama National University in engineering in 2015. }

\profile{Kimio Kuramitsu}{is an Associate Professor, leading the Language Engineering research group at Yokohama National University. His research interests range from programming language design, software engineering to data engineering, ubiquitous and dependable computing. He has received the Yamashita Memorial Research Award at IPSJ. His pedagogical achievements include Konoha and Aspen, the first programming exercise environment for novice students. He earned his B. E. at the University of Tokyo, and his Ph. D. at the University of Tokyo under the supervision of Prof. Ken Sakamura.}

\end{biography}

\end{document}